\documentclass[showpacs,showkeys,prl,twocolumn]{revtex4}%
\usepackage{amsfonts}
\usepackage{amsmath}
\usepackage{amssymb}
\usepackage{graphicx}%
\begin{document}
\title{Photon mass and quantum effects of the Aharonov-Bohm type}
\author{G. Spavieri$^{1}$ and M. Rodriguez$^{2}$}
\affiliation{$^{1}$Centro de F\'{\i}sica Fundamental, Facultad de Ciencias, Universidad de
Los Andes, M\'{e}rida, 5101-Venezuela }
\email{spavieri@ula.ve}
\affiliation{$^{2}$Departamento de F\'{\i}sica, FACYT, Universidad de Carabobo, Valencia, 2001-Venezuela}

\begin{abstract}
The magnetic field due to the photon rest mass $m_{ph}$ modifies the standard
results of the Aharonov-Bohm effect for electrons, and of other recent quantum
effects. For the effect involving a coherent superposition of beams of
particles with opposite electromagnetic properties, by means of a table-top
experiment, the limit $m_{ph}\simeq10^{-51}g$ is achievable, improving by 6
orders of magnitude that derived by Boulware and Deser for the Aharonov-Bohm effect.

\end{abstract}

\maketitle

\section{Introduction}

The possibility that the photon possesses a finite mass and its physical
implications have been discussed theoretically and investigated experimentally
by several researchers \cite{br}-\cite{lu}. Originally, the finite photon mass
$m_{\gamma}$ (measured in \textit{centimeters}$^{-1}$) has been related to the
range of validity of the Coulomb law \cite{br}, \cite{wfh}. If $m_{\gamma}%
\neq0$ this law is modified by the Yukawa potential $U(r)=e^{-m_{\gamma}%
\,r}/r$, with $m_{\gamma}^{-1}=\hbar/m_{ph}c=\lambda_{C}/2\pi$ where $m_{ph}$
is expressed in \textit{grams} and $\lambda_{C}$ is the Compton wavelength of
the photon.

There are direct and indirect tests for the photon mass. The best result
obtained so far through developments of the original Cavendish technique for
direct tests of Coulomb's law is still that from 1971 by Williams, Faller and
Hill \cite{wfh}. The null result of this experiment expressed in the form of
range of the photon rest mass is $m_{\gamma}^{-1}>3\times10^{9}cm$. Indirect
methods are provided by geomagnetic and astronomical tests\textit{.} By
studying the behavior of the magnetic field of planets, Davis, Goldhaber and
Nieto \cite{dgn} were able to determine that $m_{\gamma}^{-1}>5\times
10^{10}cm$.

Other indirect verifications are related to lumped circuit tests \cite{fa},
cryogenic experiments \cite{raa}, and the ambient\ cosmic vector
potential\textit{ }method developed by Lakes \cite{la}. If $m_{\gamma}\neq0$
the ambient\ cosmic vector potential acquires physical significance and may
interact with the dipole field of a magnetized toroid. The related experiment
by Luo, Tu, Hu, and Luan \cite{lu} yielded the range $m_{\gamma}%
^{-1}>1.66\times10^{13}cm$ and corresponding photon mass $m_{ph}<$
$2.1\times10^{-51}g$.

Several conjectures related to the Aharonov-Bohm (AB) effect \cite{ab} have
been developed assuming electromagnetic interaction of fields of infinite
range, i.e., zero photon mass. The possibility that any associated effects
become manifest within the context of finite-range electrodynamics has been
discussed by Boulware and Deser (BD) \cite{bd}. In this paper we consider and
extend BD's approach, and evaluate the limits of the photon mass that can be
determined by means of recent, new quantum effects of the Aharonov-Bohm type.

\section{Photon mass and new effects of the Aharonov-Bohm type}

In their approach, BD consider the coupling of the photon mass $m_{\gamma}$,
as predicted by the Proca equation, and calculate the resulting magnetic field
$\mathbf{B}$
\begin{equation}
\partial_{\nu}F^{\mu\nu}+m_{\gamma}^{2}A^{\mu}=J^{\mu},\;\;\;\;\mathbf{B}%
=\mathbf{B}_{0}+\widehat{\mathbf{k}}\,m_{\gamma}^{2}\,\Pi(\rho), \label{pe}%
\end{equation}
that might be used in a test of the AB effect. The first term, $\mathbf{B}%
_{0}$, is the standard magnetic field for zero photon mass --- the field
confined inside a long solenoid of radius $a$ and carrying the current $j$ ---
and the second term $\Delta\mathbf{B}=\widehat{\mathbf{k}}\,m_{\gamma}%
^{2}\,\Pi(\rho)$ represents a correction due to the photon nonvanishing rest
mass $m_{\gamma}$. Because of the extra mass-dependent term, BD obtained a
nontrivial limit on the range of the transverse photon from a table-top
experiment: $m_{\gamma}^{-1}>1.4\times10^{7}cm$. In the case of the standard
Aharonov-Casher (AC) effect \cite{AC}, this analysis has been performed by
Fuchs \cite{f} who points out that, for a neutral particle with a magnetic
dipole moment that couples to nongauge fields, no observable corrections are expected.

After the AB and AC effects, other quantum effects of this type have been
developed. Thus, it would be interesting to consider other effects of the AB
type, such as those associated with neutral particles that have an intrinsic
magnetic \cite{sdual} or electric dipole moment \cite{s}-\cite{T}, and those
with particles possessing opposite electromagnetic properties, such as
opposite dipole moments or charges \cite{s}, \cite{sep}-\cite{dow}. In the
next Sections we consider the impact of some of these new effects on the
photon mass.\ The goal would be to see if they provide similar correction
terms that might be suitable for setting more precise limits on the range of
$m_{\gamma}^{-1}$.

Before dealing with effects for electric dipoles, we recall that in Eq.
(\ref{pe}) the quantity $\Pi(\rho)$ can be expressed in terms of the Bessel
functions $I_{0}(m_{\gamma}\rho)$ and $K_{0}(m_{\gamma}\rho)$, which are
regular at the origin and infinity respectively, and reads%
\begin{align*}
\Pi(\rho)  &  =j\theta(a-\rho)[K_{0}(m_{\gamma}\rho)\int_{0}^{\rho}%
I_{0}(m_{\gamma}\rho^{\prime})\rho^{\prime}d\rho^{\prime}\\
&  +I_{0}(m_{\gamma}\rho)\int_{\rho}^{a}K_{0}(m_{\gamma}\rho^{\prime}%
)\rho^{\prime}d\rho^{\prime}]\\
&  -j\theta(\rho-a)K_{0}(m_{\gamma}\rho)\,\int_{0}^{a}I_{0}(m_{\gamma}%
\rho^{\prime})\rho^{\prime}d\rho^{\prime}.
\end{align*}

\section{Effects for electric dipoles}

The interaction term of all the effects for electric dipoles has the same
strength \cite{s}-\cite{T} so that, for the purpose of performing a table-top
experiment, we find it convenient to analyze the Tkachuk effect \cite{T}
because the resulting equations for the mass correction possess a symmetry
analogous to that of the AB effect.

For the Tkachuk effect \cite{T} we can consider a long solenoid with the
magnetization linear density $\mu=\overline{\mu}z$ and a magnetic flux
$\Phi=BS=4\pi\overline{\mu}z=\pi\overline{j}z\,a^{2}$, where $a$ is the radius
of the solenoid and $\overline{j}z$ its current density. The resulting vector
potential reads $\mathbf{A=A}_{AB}\,z$, where $\mathbf{A}_{AB}$ is the vector
potential of the AB effect with $\mu_{AB}$ substituted by $\overline{\mu}$.

The starting equation is $(-\mathbf{\nabla}^{2}+m_{\gamma}^{2})\mathbf{A}%
=\mathbf{J}$ with $\mathbf{J}=$ $(4\overline{\mu}/a^{2})z\hat{\mathbf{\phi}%
}\delta(\rho-a)$. The only difference with the AB effect is that the current
depends on $z$. Separation of variables with $\mathbf{A}=z\,\mathbf{A}%
_{T}(x,y)$ yields $(-\mathbf{\nabla}^{2}+m_{\gamma}^{2})\mathbf{A}_{T}%
(\rho,\phi)=4(\overline{\mu}/a^{2})\hat{\mathbf{\phi}}\delta(\rho-a){}$, which
is the same equation of BD. Thus, the mass treatment for the Tkachuk effect
for the electric dipole $\mathbf{d}=d\,\hat{\mathbf{k}}$ can be reduced to
that of BD.

The magnetic field is $\mathbf{B}=\mathbf{\nabla\times A}=z\mathbf{\nabla
\times A}_{T}-\mathbf{A}_{T}\mathbf{\times\nabla}z$. In the plane of motion of
the dipole, $z=0$, and the Tkachuk phase shift is \cite{T}%
\begin{align*}
\Delta\varphi &  \propto\oint\mathbf{B\times d}\cdot d\mathbf{\ell}%
=-\oint(\mathbf{A}_{T}\mathbf{\times k)\times\mathbf{d}}\cdot d\mathbf{\ell}\\
&  =d\oint(\mathbf{A}_{T})\cdot d\mathbf{\ell}=d\int_{S}\mathbf{\nabla\times
A}_{T}\cdot d\mathbf{S}%
\end{align*}
where the last integral is the flux through the surface as in the AB effect
and BD approach. From Eq. (\ref{pe}), we write for the photon mass
contribution $\Delta\mathbf{B}(\overline{j},\rho)=\mathbf{\nabla\times
A}_{Tm_{\gamma}}(\overline{j},\rho)=\hat{\mathbf{k}}\,m_{\gamma}^{2}%
\,\Pi(\overline{j},\rho)$ so that the mass correction to the phase reads%
\[
\Delta\varphi=2\pi(d/\hbar c)\int_{a}^{\rho}\left[  (m_{\gamma}^{2}%
\,\Pi(\overline{j},\rho))\right]  \rho d\rho.
\]
In the exterior ($\rho>a$) region, $\Delta B=m_{\gamma}^{2}\,\Pi(\overline
{j},\rho))\simeq(\overline{j}/2)(m_{\gamma}a)^{2}\ln(2/m_{\gamma}\rho)$
\cite{bd}. With $4\overline{\mu}=\overline{j}a^{2}$, and the Takchuk phase
$\varphi_{0}=4\pi d\overline{\mu}/\hbar c$, the relative variation of the
phase due to the photon mass is%
\begin{equation}
\frac{\Delta\varphi}{\varphi_{0}}=\frac{\overline{j}a^{2}}{4\overline{\mu}%
}\int_{a}^{\rho}m_{\gamma}^{2}\ln(\frac{2}{m_{\gamma}\rho})\rho d\rho\sim
\frac{1}{2}(m_{\gamma}\rho)^{2}\ln(\frac{2}{m_{\gamma}\rho}). \label{dft}%
\end{equation}

Following BD \cite{bd} we set $\Delta\varphi\geq2\pi\varepsilon=2\pi
\times10^{-3}$, where $\varepsilon$ is the precision of the measurement, and
write Eq.(\ref{dft}) as%
\[
2\pi\varepsilon/\varphi_{0}=(1/2)(m_{\gamma}\rho)^{2}\ln(2/m_{\gamma}\rho).
\]
This result, valid for the Tkachuk effect, can be compared with that of BD
derived for the AB effect,%
\[
2\pi\varepsilon/\varphi_{0AB}=(1/2)(m_{\gamma BD}\rho)^{2}\ln(2/m_{\gamma
BD}\rho),
\]
where $m_{\gamma BD}$ is the value of the photon mass obtained by BD and
$\varphi_{0AB}$ is the value of the AB phase shift when $m_{\gamma}=0$. In
this case, the contribution due to the logarithmic terms is not relevant and
can be neglected. For the comparison, we use $d=e$ $a_{0}$ for the dipole with
$a_{0}$ the Bohr radius, $\overline{\mu}=\mu_{AB}/l$ with $l\simeq1cm$ the
realistic length of the solenoid in the Tkachuk effect \cite{T}, and obtain%
\[
m_{\gamma}^{-1}=m_{\gamma BD}^{-1}\left[  \frac{\varphi_{0}}{\varphi_{0AB}%
}\right]  ^{1/2}=m_{\gamma BD}^{-1}\left[  \frac{a_{0}}{l}\right]  ^{1/2}%
\sim10^{-4}m_{\gamma BD}^{-1},
\]
which represents a range limit of the photon mass 4 orders of magnitude lower
than that of BD.

As expected, no improvement for the range $m_{\gamma}^{-1}$ is achieved from a
table-top experiment involving electric dipoles because of the lower strength
of the em interaction.

\section{Effect for superposition of $\pm$ charged particles}

One of us \cite{sep} has pointed out that the observable quantity in the AB
effect is actually the phase difference
\begin{equation}
\Delta\varphi=\frac{e}{\hbar c}[\int\mathbf{A}\cdot d\mathbf{\ell-}%
\int\mathbf{A_{0}}\cdot d\mathbf{\ell]} \label{nph}%
\end{equation}
where the integral can be taken over an open path integral. For the usual
closed path $c$ encircling the solenoid and limiting the surface $S$, the
observable quantity is the phase-shift variation, $\Delta\phi\propto\oint
_{c}\mathbf{A}\cdot d\mathbf{\ell-}\oint_{c}\mathbf{A_{0}\cdot}d\mathbf{\ell
}=\oint_{S}\mathbf{B}\cdot d\mathbf{S-}\oint_{S}\mathbf{B}_{0}\cdot
d\mathbf{S}$. In fact, in interferometric experiments involving the AB and AC
effects \cite{cham}, \cite{ton}, \cite{sang}, the direct measurement of the
phase $\varphi\propto\int\mathbf{A}\cdot d\mathbf{\ell}$ or phase shift
$\phi\propto\oint\mathbf{A}\cdot d\mathbf{\ell}$ is impossible in principle
without the comparison of the actual interference pattern with an interference
reference pattern. Thus, $\varphi$ or $\phi$ are not observable, but the
variations $\Delta\varphi$ and $\Delta\phi$ are both gauge-invariant
observable quantities \cite{sep}.

It follows that, in analogy with the AC effect for a coherent superposition of
beams of magnetic dipoles of opposite magnetic moments $\pm\mu$ \cite{sang}
and the effect for electric dipoles of opposite moments $\pm d$ \cite{dow}, an
effect of the AB type for a coherent superposition of beams of charged
particles with opposite charge state $\pm q$ is theoretically feasible
\cite{sep}. In the mentioned cases, the beam of particles possessing opposite
em properties do not encircle the singularity (e. g., solenoid for the AB
effect, and line of charges for the AC effect) but travel at one side of it
along a straight path $C$. Depending on the interferometric technique used
\cite{sang}, \cite{dow}, the length of $C$ can be of the order of a few $cm$
up to a few $m$. In the experiment by Sangster \textit{et al.} \cite{sang},
the beam splitter of the magnetic dipoles $\pm\mu$ is the magnetic field
$\mathbf{B}$ of a Ramsey loop \cite{sang}. Similarly, in the experimental set
up considered by Dowling \textit{et al.} \cite{dow}, the electric dipoles $\pm
d$ are split by an electric field $\mathbf{E}$. An external uniform electric
potential $V$ could act as a possible beam splitter for particles of opposite
charge $\pm q$.

Although the effect for $\pm q$ charged particles is viable \cite{sep}, the
technology and interferometry for the test of this effect needs improvements.
It is worth recalling that not long ago the technology and interferometry for
beams of particles with opposite magnetic $\pm\mu$ or electric $\pm d$ dipole
moments was likewise unavailable, but is today a reality \cite{sang},
\cite{dow}. Discussions on this subject may act as a stimulating catalyst for
further studies and technological advances that will lead to the experimental
test of this quantum effect. An important step in this direction has already
been made \cite{sep} by showing that, at least in principle and as far as
gauge invariance requirements are concerned, this effect is physically
feasible. Therefore, using this effect in a table-top experiment analogous to
that of BD, one is entitled to ask what would be its relevance in eventually
determining a bound for the photon mass $m_{ph}$.

\subsection{Determining the mass correction $\Delta\varphi$ in the effect for
$\pm$ charged particles}

In the experimental set ups detecting the traditional AB effect there are
limitations imposed by the suitable type of interferometer related to the
electron wavelength, the corresponding convenient size of the solenoid or
toroid, and the maximum achievable size $\rho$ of the coherent electron beam
encircling the magnetic flux \cite{bd}. In the analysis made by BD, the radius
of the solenoid is $a=0.1cm$, and $\rho$ is taken to be about $10$ $cm$,
implying that the electron beam keeps its state of coherence up to a size
$\rho=10^{2}a$, i. e., fifty times the solenoid diameter. The advantage of the
new approach for the $\pm q$ beam of particles is that the dimension of the
solenoid has no upper limits and is conditioned only by practical limits of
the experimental set up, while the size of the coherent beam of particles
plays no important role.

In order to calculate the line integral appearing in Eq. (\ref{nph}) we need
the analytical expression of $\mathbf{A}(\mathbf{x})$. This can be obtained
from Stokes' theorem, $%
{\displaystyle\int}
\mathbf{B}\cdot d\mathbf{S=}%
{\displaystyle\oint}
\mathbf{A}\cdot d\mathbf{\ell}=2\pi A_{\varphi}\rho$ and solving for
$A_{\varphi}$. Another approach consists of calculating $\mathbf{A}$ from the
expression $\mathbf{A}=\widehat{\mathbf{k}}\times\mathbf{\nabla\,}\Pi\left(
\rho\right)  $ in Ref. \cite{bd}, using the recurrence relations for the
modified Bessel functions. The same result can be obtained using the approach
\cite{s}, \cite{sn} that consists of calculating the interaction
electromagnetic momentum, which in the Coulomb gauge yields $e\mathbf{A}/c$.
The result is%
\[
A_{\varphi}=j\frac{a^{2}}{2}\dfrac{1}{\rho}+\left(  \frac{j}{2}\right)
\left(  m_{\gamma}a\right)  ^{2}\frac{\rho}{2}\ln\left(  \frac{m_{\gamma}\rho
}{2}\right)  .
\]
Taking the path $C$ along the $x$ axis for a path length $2x$ with $x>>y$ we
find $\varphi_{0}=\int_{C}\mathbf{A}_{m_{\gamma}=0}\cdot d\mathbf{\ell}%
\simeq-(\pi/2)a^{2}j$. The contribution due to $m_{\gamma}$ yields%
\[
\int_{C}\mathbf{A}\cdot d\mathbf{\ell}=\left(  j/2\right)  \left(  m_{\gamma
}a\right)  ^{2}yx\ln\left(  m_{\gamma}\sqrt{x^{2}+y^{2}}/2\right)
\]
for the same path length $2x$. Consequently, from Eq. (\ref{nph}) and Ref.
\cite{sep} the observable phase shift variation is $\Delta\varphi=2j\left(
m_{\gamma}a\right)  ^{2}yx\ln\left(  m_{\gamma}\sqrt{x^{2}+y^{2}}/2\right)  $
and%
\begin{equation}
\frac{\Delta\varphi}{\varphi_{0}}=-\frac{4}{\pi}m_{\gamma}^{2}\,xy\ln\left(
m_{\gamma}\sqrt{x^{2}+y^{2}}/2\right)  . \label{dfn}%
\end{equation}

\subsection{Evaluating the photon mass limit}

Following BD \cite{bd} we set $\Delta\varphi\geq2\pi\varepsilon=2\pi
\times10^{-3}$ where $\varepsilon$ is the precision of the measurement. The
value of $m_{\gamma}$ at which the effect is just observable is%
\[
\frac{2\pi\varepsilon}{\varphi_{0}}=-\frac{4}{\pi}m_{\gamma}^{2}\,xy\ln\left(
m_{\gamma}\sqrt{x^{2}+y^{2}}/2\right)  .
\]

This value can be compared with the corresponding one by BD \cite{bd}, while,
as done by BD, we neglect the small corrections due to the contribution of the logarithms.

The question is now: what would be the size of the solenoid in order to
achieve a photon mass limit of the order of that of Ref. \cite{lu} found by
Luo \textit{et al.}? We estimate $m_{\gamma}$ with respect to $m_{\gamma BD}$
for an ideal experimental set up that, apart from considerations of cost, is
realistically within reach of present technology. For a vector potential
produced by the magnet of a huge cyclotron-type solenoid (radius $a=5m$ and
length or height $D$ several times the radius), we estimate $\varphi
_{0}/\varphi_{0BD}$ $\simeq a^{2}/(a_{BD})^{2}=5^{2}/(10^{-3})^{2}$. For a
path of $x=6a=300\rho$ at the distance $y=80\rho$ we obtain%
\begin{align}
m_{\gamma}^{-1}  &  =m_{\gamma BD}^{-1}\left[  \frac{8}{\pi}\frac{\varphi_{0}%
}{\varphi_{0BD}}\frac{xy}{\rho^{2}}\right]  ^{1/2}\simeq10^{6}\,m_{\gamma
BD}^{-1}\label{fr}\\
&  \simeq2\times10^{13}cm.\nonumber
\end{align}
With their table-top experiment, BD obtained the value $m_{\gamma BD}%
^{-1}\simeq140Km$ that is equivalent to $m_{phBD}=2.5\times10^{-45}g$. With
our approach, the new limit (\ref{fr}) of the photon mass is $m_{ph}%
\simeq2\times10^{-51}g$ which is of the same order of magnitude of that found
by Luo \textit{et al.} \cite{lu}.

\subsection{Secondary effects}

When the AB effect was tested for the first time \cite{cham}, physicists were
concerned about the effect of the stray fields just outside the solenoid on
the phase shift of electrons that was going to be observed. The stray field
$\Delta\mathbf{B}$ acts on the beam of charges, bends it, and displaces the
interference pattern. Depending on the technique used for observing the AB
phase shift, the effect of the stray field may mask, or not, the AB phase
shift. With the approach used by Chambers \cite{cham}, the interference
pattern is shifted by a large amount but the figure of the pattern is left
unaltered. However, the AB effect changes the figure of the pattern so that
the AB phase shift is easily observable. In this case the two effects, that of
the stray fields and of the AB phase shift, can be separated and observed. The
field $\Delta\mathbf{B}=m_{\gamma}^{2}\Pi(\rho)\widehat{\mathbf{k}}$ produces
a variation of the particle momentum $\delta p_{\perp}\propto%
{\displaystyle\int}
ev\Delta Bdt$ in the direction perpendicular to the direction of motion
$\mathbf{v}$. The corresponding angular deflection of the beam, $\alpha
\simeq\delta p_{\perp}/p$, can be estimated and the equivalent shift of the
interference pattern, considering a beam of particles through a double-slit
interferometer, can be determined. The equivalent resulting phase shift
$(\Delta\varphi)_{\Delta B}$ is smaller by about two orders of magnitude than
the phase shift value $\Delta\varphi_{BD}$ found by BD \cite{bd}.

Although this $(\Delta\varphi)_{\Delta B}$ is small and does not lead to
important phase shift variations with respect to $\Delta\varphi_{BD}$, the
fact that the leakage field $\Delta\mathbf{B}=m_{\gamma}^{2}\Pi(\rho
)\widehat{\mathbf{k}}$ bends the electron beam suggests that, within a
classical approach, $m_{\gamma}$ may be estimated by measuring directly the
angular or linear deflection of the beam. In order to magnify the deflection
and improve the $m_{\gamma}^{-1}$ range, it would be convenient to use the
magnetic field generated by the cyclotron-type solenoid, as described above,
supposing ideally that the other stray fields due to imperfections in the
construction of the solenoid and to its finite length can be taken into
account separately.

The effect of the leakage field $\Delta\mathbf{B}=m_{\gamma}^{2}\Pi
(\rho)\widehat{\mathbf{k}}$ will not be considered in this paper and will be
discussed in detail elsewhere. We simply mention that measurements of the
linear displacement $s_{\perp}\simeq\delta s_{\perp}$ performed with this
approach, that treats the electron as a classical particle, possesses quantum
restrictions and can be meaningful only up to the value of the bound
established by Heisenberg's uncertainty principle $\delta p_{\perp}\delta
s_{\perp}\simeq h$.

\section{Conclusions}

We have considered the table-top approach of BD and extended it to several
effects of the AB type. In discussing the quantum effects for electric dipoles
and comparing them with the BD approach, we have found no improvement for the
$m_{\gamma}^{-1}$ range in this case. For the case of the AB solenoid,
improvements are possible by taking into account the effect of the leakage
field $\Delta\mathbf{B}$ on the beam of particles, testing its effect on the
bending of the beam in a classical approach.

Moreover, if a cyclotron-type solenoid is used for testing the $\pm
q$\ quantum effect proposed by Spavieri \cite{sep}, a photon mass bound of the
value of $m_{ph}\sim10^{-51}g$ should be achievable. This result represents a
lower limit improvement of 6 orders of magnitude with respect to the approach
of BD with the standard AB effect. The latest results by Luo \textit{et al.}
\cite{lu} and the prospects of the AB type of quantum effect scenario here
discussed are certainly remarkable if one considers that, according to the
uncertainty principle, a purely theoretical estimate of the photon mass is
given by $m_{ph}=h/(\Delta t)c^{2}$ which yields an order of magnitude number
of $m_{ph}=10^{-65}g$, where the age of the universe is taken to be roughly
$10^{10\text{ }}$years.

In closing, advances in the area related to the AB type of effects indicate
that the photon mass limit achievable with this quantum approach could compete
with other methods. However, it is not only a question of improving the
limits, but of extending the scenario where tests of the photon mass can be
realized, as in the cryogenic photon-mass experiment performed by Ryan
\textit{et al.} \cite{raa} where the validity of the results is extended from
the standard terrestrial (`room') temperatures to those of the galactic
environment. Each approach is important in itself as it extends the range of
validity of the Coulomb law and of the $m_{\gamma}^{-1}$ range as a function
of the physical conditions of the measurement.

\section{Acknowledgments}

This work was supported in part by the CDCHT (Project C-1413-06-05-A), ULA,
M\'{e}rida, Venezuela.

\end{document}